\documentclass[twocolumn]{aastex631}
\usepackage{amsmath}
\usepackage{natbib}
\usepackage{graphicx}
\usepackage{url}
\usepackage{soul}

\newcommand{\rinbest}{96}
\newcommand{\deltarbest}{1}
\newcommand{\halfanglebest}{5}
\newcommand{\massbest}{4.5}
\newcommand{\aminbest}{1.4}

\newcommand{\ringood}{92--100}
\newcommand{\deltargood}{$\leq$10}
\newcommand{\halfanglegood}{$\geq$1.4}
\newcommand{\massgood}{4--5}
\newcommand{\amingood}{1--2}

\newcommand{\chisqthresh}{1.2}
\newcommand{\numgoodmodels}{41}
\newcommand{\nummodelsingrid}{9408}

\newcommand{\Hinmin}{$\geq$ 2.2}

\newcommand{\lifetimesec}{2.8 $\times$ $10^9$}
\newcommand{\lifetimeyr}{87}
\newcommand{\Mdeq}{1.0 $\times$ 10$^{18}$}

\usepackage{verbatim}

\shorttitle{Modeling the G29-38 Disk}
\shortauthors{Ballering et al.}

\begin{document}

\title{The Geometry of the G29-38 White Dwarf Dust Disk from Radiative Transfer Modeling}

\author[0000-0002-4276-3730]{Nicholas P. Ballering}
\altaffiliation{Virginia Initiative on Cosmic Origins Fellow.}
\affiliation{Department of Astronomy, University of Virginia, Charlottesville, VA 22904, USA}

\author{Colette I. Levens}
\affiliation{Department of Atmospheric, Oceanic and Planetary Physics, University of Oxford, Clarendon Laboratory, Oxford, OX1 3PU, UK}

\author[0000-0002-3532-5580]{Kate Y. L. Su}
\affiliation{Steward Observatory, University of Arizona, 933 North Cherry Avenue, Tucson, AZ 85721, USA}

\author[0000-0003-2076-8001]{L. Ilsedore Cleeves}
\affiliation{Department of Astronomy, University of Virginia, Charlottesville, VA 22904, USA}
\affiliation{Department of Chemistry, University of Virginia, Charlottesville, VA 22904, USA}

\correspondingauthor{Nicholas P. Ballering}
\email{nb2ke@virginia.edu}

\begin{abstract}
Many white dwarfs host disks of dust produced by disintegrating planetesimals and revealed by infrared excesses. The disk around G29-38 was the first to be discovered and is now well-observed, yet we lack a cohesive picture of its geometry and dust properties. Here we model the G29-38 disk for the first time using radiative transfer calculations that account for radial and vertical temperature and optical depth gradients. We arrive at a set of models that can match the available infrared measurements well, although they overpredict the width of the 10 $\micron$ silicate feature. The resulting set of models has a disk inner edge located at \ringood\,$R_\text{WD}$ (where $R_\text{WD}$ is the white dwarf radius). This is farther from the star than inferred by previous modeling efforts due to the presence of a directly illuminated front edge to the disk. The radial width of the disk is narrow (\deltargood\,$R_\text{WD}$); such a feature could be explained by inefficient spreading or the proximity of the tidal disruption radius to the sublimation radius. The models have a half-opening angle of \halfanglegood$^\circ$. Such structure would be in strong contradiction with the commonly employed flat-disk model analogous to the rings of Saturn, and in line with the vertical structure of main-sequence debris disks. Our results are consistent with the idea that disks are collisionally active and continuously fed with new material, rather than evolving passively after the disintegration of a single planetesimal.
\end{abstract}

\section{INTRODUCTION}
\label{sec:introduction}

Dust disks around white dwarfs are evident from their infrared (IR) excess, and dozens of such disks have been discovered \citep[e.g.][]{rebassa-mansergas2019,dennihy2020_WISE,xu2020,lai2021}. The dust likely originates from planetesimals scattered into highly eccentric orbits then tidally disrupted upon close approach to the white dwarf star \citep{debes2002,jura2003,veras2014_WDdiskformI}. Actively disintegrating planetesimals have been seen via their transit signatures in a growing number of systems \citep{vanderburg2015_transits,vanderbosch2020_transits,vanderbosch2021_transits,guidry2021_transits,farihi2022_transits}
although the process by which the resulting debris forms a dust disk remains an active area of investigation \citep[e.g.,][]{veras2015_WDdiskformII,brouwers2022_WDdisk}.

Disk material is fed onto the star itself and observed as metal-rich pollution in the white dwarf photosphere. Hundreds of polluted white dwarfs are now known. Analysis of polluted white dwarfs allows the elemental abundance of the original planetesimals to be inferred \citep[e.g.,][]{xu2019_WDcomp}. Gas disks have also been found around 21 white dwarfs, seen via emission or absorption lines of atomic Ca, Fe, O, Mg, Si, and Na \citep[e.g.,][]{gansicke2006,dennihy2020_gas,melis2020,gentile-fusillo2021}. How the dust and gas disks interact, why gas disks are less common than dust disks, and the mechanism by which material is accreted onto the star are also areas of ongoing investigation \citep[e.g.,][]{rafikov2011_WDaccretion,metzger2012_WDaccretion,swan2020_WDvariability,swan2021_WDcollisions,li2021_WDaccretion}. See \citet{farihi2016} and \citet{veras2021_WDreview} for recent reviews of white dwarf disks and pollution.

G29-38---the subject of this study---was the first white dwarf with a detected IR excess \citep{zuckerman1987,graham1990_G2938,tokunaga1990_G2938}, and it remains one of the brightest and most well-observed dust disks. As a result, the system has a wealth of data from broadband photometry to spectroscopy, covering wavelengths from the near-IR to the submillimeter. Various models have been constructed to fit the available data. An early model was geometrically flat, analogous to the rings of Saturn, as expected from collisional damping of the dust's vertical motion \citep{jura2003}. This model is optically thick, resulting in featureless IR emission that depends on the temperature profile between inner and outer disk radii and inclination. The flat-disk model fit the available photometry at the time, and it is still employed for many systems with limited data \citep[e.g.][]{gentile-fusillo2021}. However, Spitzer/Infrared Spectrograph (IRS) spectroscopy revealed mineralogical features, including a prominent silicate emission feature at $\sim$10 $\micron$, which are inconsistent with entirely optically thick emission \citep{reach2005_G2938}. Several other bright disks with IRS spectra also exhibit mineralogical features \citep{jura2007,jura2009_wdsilicates}, suggesting that, aside from being brighter, the G29-38 disk may not be atypical.

Other models of the G29-38 disk have accounted for an optically thin component in order to fit the mineralogical features. Purely optically thin models were used by \citet{reach2005_G2938,reach2009}, with the latter study conducting a detailed compositional analysis of the dust. Their model did not explicitly prescribe the dust location or disk geometry, but for a reasonable dust distribution the inferred mass would not be entirely optically thin, so the model was not fully self-consistent. \citet{reach2009} also explored moderately optically thick models and models combining optically thick and thin components. The optically thick component was posited as a flat disk and the thin component either a diffuse halo or warped outer disk, following the model setup \citet{jura2007} used to interpret the GD 362 dust disk. \citet{xu2018} also used a two-component (optically thin plus optically thick) model to fit the G29-38 IR spectral energy distribution (SED) and interpret the observed variability. None of these two-component models represented a complete and self-consistent picture of the dust disk.

The goal of this paper is to forward model the dust emission around G29-38 with a cohesive disk structure similar to a protoplanetary disk or main-sequence debris disk. IR spectra of the model are generated using a radiative transfer code that naturally accounts for radial and vertical temperature and optical depth gradients in the disk. We describe the G29-38 data and how we fit them with radiative transfer models in Section \ref{sec:methods}. In Section \ref{sec:results} we present the model fits, illustrate the effect of each model parameter on the simulated spectra, and show our derived model dust temperature. In Section \ref{sec:discussion} we discuss the implications of the disk's grain sizes, mass, vertical structure, and radial structure, and we compare the results with the general properties of main-sequence debris disks. We also discuss the limitations of our models. We summarize our results in Section \ref{sec:summary}.

\section{Methods}
\label{sec:methods}

In short, our methodology entails compiling photometry and spectra of the G29-38 system from the literature and fitting these data with spectra of a model disk. The model geometry is parameterized to resemble a protoplanetary disk or main-sequence debris disk. We use the RADMC-3D radiative transfer tool \citep{dullemond2012} to generate the model spectra. We constrain the disk parameters by fitting a grid of models to the data.

\subsection{Archival Data Collection}

We use two sets of IR spectra from the Spitzer Space Telescope \citep{werner2004_spitzer} IRS \citep{houck2004_IRS}: AOR 13828096 from 2006 (SL2, LL2, and LL1 modules covering 5.2--8.7 and 14.0--38 $\micron$) and AOR 22957568 from 2007 (SL2 and SL1 modules covering 5.2--14.5 $\micron$). These spectra have been previously presented in the literature \citep{reach2009,farihi2014_G2938,xu2018}. We obtain the spectra from the Combined Atlas of Sources with Spitzer IRS Spectra \citep{lebouteiller2011_CASSIS} online database. The IRS data have uncertainties of $\sim$5\% at $\lambda$ $<$ 12 $\micron$ and larger uncertainties at longer wavelengths.  We exclude the measurements at wavelengths longer than 30 $\micron$ because they have a very low signal-to-noise ratio. Another IRS spectrum covering 5.2--14.5 $\micron$ was acquired in 2004 (AOR 10184192), which we do not use in our analysis. It showed a 10\% weaker excess than the 2007 spectrum, and this difference was analyzed by \citet{xu2018}.    

We also use spectra from the IRTF/SpeX instrument \citep{rayner2003_Spex} originally published by \citet{kilic2006_WDs}. The spectra were provided by M. Kilic (private communication). We use only the data from 1.3--2.5 $\micron$, as shorter wavelengths trace the white dwarf photosphere, not the disk. We re-scale this spectrum by a factor of 0.96 so the short end aligns with our white dwarf photosphere model, as was also found to be necessary by \citet{reach2009}. We assign uncertainty values of 5\% to the points in this spectrum.  

Finally, we compile IR photometry from the literature (Table \ref{tab:photometry}). G29-38 is known to be variable in the optical and near-IR \citep{graham1990_G2938,kleinman1998_G2938} due to stellar pulsation. This can lead to systematic offsets between measurements that are not taken simultaneously and likely explains why the J- and H-band photometry are somewhat fainter than our model photosphere (see Section \ref{sec:bestfit}). The IRAC photometry was measured over multiple epochs \citep{xu2018}; we adopt the median value and use the standard deviation of the measurements for the uncertainty. No uncertainty estimates were provided for the IRS peakup photometry \citep{reach2005_G2938}, so we adopt a conservative estimate of 10\%. G29-38 was not detected at 70 $\micron$ with Spitzer/MIPS \citep{jura2009_XrayMIPS}, at 100 or 160 $\micron$ with Herschel/PACS, nor in the (sub)mm with ALMA \citep{farihi2014_G2938}. We do not formally include these measurements in our fitting process, but we note that our models that fit the IR detections are also consistent with these upper limits.

\begin{deluxetable}{lccc}
\tablecaption{Photometric Data \label{tab:photometry}}
\tablewidth{0pt}
\tablehead{
\colhead{$\lambda$ ($\micron$)} & \colhead{$F_\nu$ (mJy)} & \colhead{Instrument} & \colhead{Reference}}
\startdata
1.25 &   9.00  $\pm$  0.24   &   2MASS & \citet{cutri2003_2MASScat} \\
1.63 &   6.15  $\pm$  0.24    &  2MASS & \citet{cutri2003_2MASScat} \\
2.19 &   5.55  $\pm$  0.16    &  2MASS & \citet{cutri2003_2MASScat} \\
3.35 &   7.68  $\pm$  0.17    &  WISE  & \citet{cutri2012_WISEcat} \\
3.55 &   8.08 $\pm$  0.15  &  IRAC  & \citet{xu2018} \\
4.49 &   8.99 $\pm$  0.19   &  IRAC  & \citet{xu2018} \\
4.60 &   8.73  $\pm$  0.17    &  WISE  & \citet{cutri2012_WISEcat} \\
5.73 &   8.27 $\pm$  0.11  &  IRAC  & \citet{xu2018} \\
7.87 &   8.48  $\pm$  0.24  &  IRAC  & \citet{xu2018} \\
10.50 &  11.1 $\pm$  2.2     &  IRTF  & \citet{tokunaga1990_G2938} \\
11.60 &  7.39  $\pm$  0.20    &  WISE  & \citet{cutri2012_WISEcat} \\
16.0  &  3.7   $\pm$  0.37    &  IRS   & \citet{reach2005_G2938} \\
22.10 &  2.4   $\pm$  1.02    &  WISE  & \citet{cutri2012_WISEcat} \\
23.68 &  2.75  $\pm$  0.05    &  MIPS  & \citet{xu2018} \\
\enddata
\end{deluxetable}

\subsection{Disk Model}

We model the dust distribution around G29-38 as an azimuthally symmetric disk with a geometry like that of a protoplanetary disk or main-sequence debris disk. While the structures of protoplanetary and debris disks have been revealed by resolved imaging, this is not possible for white dwarf disks owing to their much smaller sizes, so white dwarf disk structure is much more uncertain. Thus, we take a simplified approach and acknowledge reality may be more complex.

The radial distribution of our dust disk model follows a power law between $r_\text{in}$ and $r_\text{out} = r_\text{in} + \Delta r$. That is, the surface density follows
\begin{equation}
    \Sigma(r) = \Sigma_\text{in} \left(\frac{r}{r_\text{in}}\right)^{-\gamma}.
\end{equation}
We fix $\gamma$ = 1, a typical value found for protoplanetary disks from resolved submillimeter images \citep{andrews2010}. We parameterize the amount of dust in terms of total mass, $M$, so the surface density at the inner edge of the disk is 
\begin{equation}
    \Sigma_\text{in} = \frac{M(2-\gamma)}{2 \pi r_\text{out}^2\left(\frac{r_\text{out}}{r_\text{in}}\right)^{-\gamma} - 2 \pi r_\text{in}^2}.
\end{equation}
The dust distribution decreases exponentially above and below the midplane, with volume density
\begin{equation}
\rho(r,z) = \frac{\Sigma(r)}{\sqrt{2\pi} H(r)} \exp\left[-\frac{1}{2}\left(\frac{z}{H(r)}\right)^2\right].
\end{equation}
The disk scale height, $H$, increases linearly with $r$, so we parameterize the vertical structure using the half-opening angle of the disk, $\Theta$, i.e., $H(r) = r \tan(\Theta)$. In protoplanetary disks, the scale height increases with radius faster than linearly; that is, the disks are flared. This is due to vertical support from the large gas reservoir in hydrostatic equilibrium. No gas is detected in the G29-38 disk, so our disk model is not flared.     

While $r$ and $z$ are cylindrical coordinates, we use a spherical coordinate grid to set up the radiative transfer simulation, as RADMC-3D does not support cylindrical coordinates. We use a regular grid with 100 cells in the radial direction, spaced logarithmically between $r_\text{in}$ and $\sqrt{r_\text{out}^2 + (4 H_\text{out})^2}$, ensuring that the full radial extent of the disk is sampled to four scale heights. We use 116 cells in the polar coordinate, $\theta$, with higher density within four scale heights of the midplane. Specifically, we use nine cells from $0 < \theta < \frac{\pi}{2} - 4 \Theta$, 98 cells from $\frac{\pi}{2} - 4 \Theta < \theta < \frac{\pi}{2} + 4 \Theta$, and another nine cells from $\frac{\pi}{2} + 4 \Theta < \theta < \pi$. The cells are linearly spaced within each region. We use 20 cells, equally spaced, in the azimuthal direction.

We use amorphous astronomical silicates for the dust composition, with optical constants from \citet{draine2003}\footnote{Accessed from  \url{https://www.astro.princeton.edu/~draine/dust/dust.diel.html}}. \citet{xu2018} used either a combination of astronomical silicates and crystalline forsterite or olivine grains, along with an opaque disk component, to produce a decent fit to the G29-38 spectra. The crystalline component contributes more subtle and narrow spectral features compared to the astronomical silicates. We ignore the crystalline component, as our goal is to fit the most prominent structure in the IR spectrum. We generate a library of opacity spectra for the dust at a range of grain sizes using their optical constants and Mie theory. We do this using a code included in the RADMC-3D software package based on the algorithm by \citet{bohren1983}.

We model a power-law distribution of grain sizes, $n(a) \propto a^{-3.5}$ between $a_\text{min}$ and $a_\text{max}$ \citep{mathis1977_MRNdist}, as is typically found for protoplanetary and debris disks. We fix $a_\text{max}$ = 5 $\micron$ but leave $a_\text{min}$ as a free parameter. To sample the distribution, we populate the RADMC-3D grid with 30 grain sizes, each treated as a separate ``species." All grain sizes follow the same spatial distribution.

The disk is heated by radiation from the central white dwarf star, which we model as a 11,240 K spherical blackbody with a radius of $R_\text{WD}$ = 9.04 $\times$ $10^8$ cm \citep{xu2018}. The radiative transfer simulation proceeds in two steps. First, the dust temperature is computed throughout the disk using $10^7$ photon packets sampling wavelengths from 0.08 to 1000 $\micron$. The radiative transfer is accelerated in high-optical-depth parts of the disk with a modified random walk procedure. Scattering is not included in the radiative transfer. Second, to compare with observations, the thermal emission spectrum of the star plus disk is calculated using $10^4$ photon packets. The spectrum is computed on a custom wavelength grid from 0.1--1500 $\micron$ sampled finely ($\Delta\lambda = 0.05\,\micron$) from 5.1--38 $\micron$ and more coarsely elsewhere. Following \citet{xu2018}, we fix the disk inclination to $i$ = 30$^\circ$, and we discuss the effect of inclination in Section \ref{sec:limitations}. We normalize the resulting spectrum to a distance of 17.5 pc.

\subsection{Fitting Procedure}

\begin{deluxetable}{lccc}
\tablecaption{Model Parameters \label{tab:parameters}}
\tablewidth{0pt}
\tablehead{
\colhead{Parameter} & \colhead{Description} & \colhead{Best Fit} & \colhead{Good Fits\tablenotemark{a}}}
\startdata
$r_\text{in}$ ($R_\text{WD}$) & Inner radius & \rinbest & \ringood \\
$\Delta r$ ($R_\text{WD}$) & Radial width & \deltarbest & \deltargood \\
$\Theta$ (deg.) & Half-opening angle & \halfanglebest & \halfanglegood \\
$M$ ($10^{18}$ g) & Dust mass & \massbest & \massgood \\
$a_\text{min}$ ($\micron$) & Minimum grain size & \aminbest & \amingood \\
\enddata
\tablenotetext{a}{Good fits are models with $\chi^2$ $<$ $\chisqthresh$ times the minimum $\chi^2$}
\end{deluxetable}

To search for the best-fit model, we vary five parameters: $r_\text{in}$, $\Delta r$, $\Theta$, $M$, and $a_\text{min}$. The following parameters remain fixed: $\gamma$ = 1, $a_\text{max}$ = 5 $\micron$, and $i$ = 30$^\circ$. We parameterize $r_\text{in}$ and $\Delta r$ in units of $R_\text{WD}$, which provides a convenient scale for the size of the disk. For reference, $100 R_\text{WD}$ = 0.006 au.

We evaluate the fit according to the $\chi^2$ metric. We first search for a good fit by exploring parameter space in an ad hoc manner. We then build a grid of models centered around our best-fit model. The grid entails $\nummodelsingrid$ models. To quantify the range of parameters that yield good fits, we examine the set of models with $\chi^2$ values within some factor of the lowest $\chi^2$. We conclude that models with $\chi^2$ $<$ $\chisqthresh$ times the minimum $\chi^2$ achieve a good fit to the data.

When computing $\chi^2$, we did not distinguish between photometric and spectroscopic data points. Thus, the spectra, due to having more data points, naturally influence the fit more so than does the photometry. Some studies that fit IR SEDs counteract this by increasing the weight given to photometry when computing $\chi^2$ \citep[e.g.,][]{ballering2013}. This may be necessary when the spectra and photometry sample non-overlapping wavelengths (e.g., mid-IR and far-IR). In this case, the spectra and photometry sample similar wavelengths, so up-weighting the photometry was not necessary.

\section{Results}
\label{sec:results}

\begin{figure}
\epsscale{1.17}
\plotone{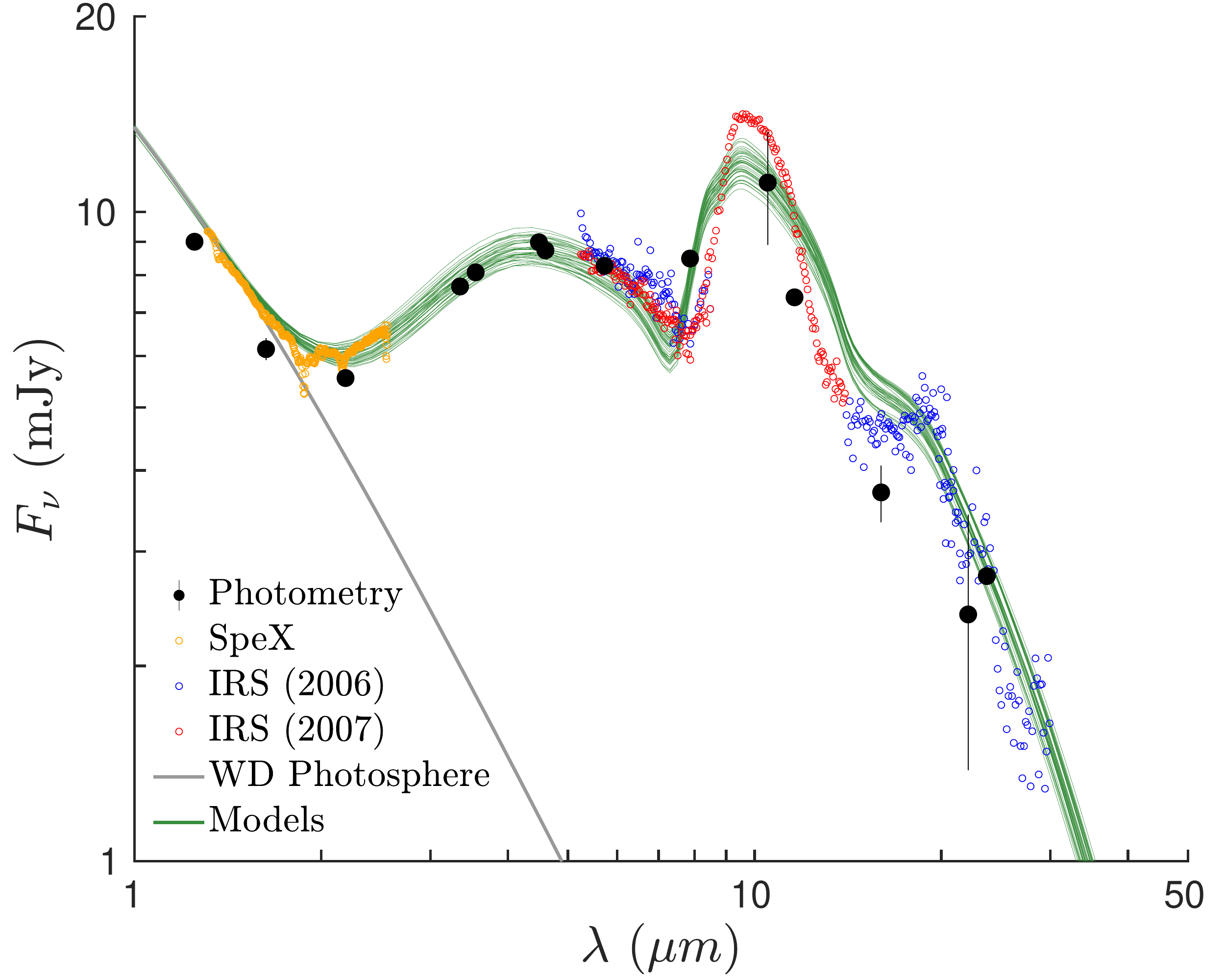}
\caption{Measured infrared spectral energy distribution of G29-38 along with the \numgoodmodels\,good-fitting models (green). For clarity, we do not show error bars on the spectroscopic data points.}
\label{fig:bestfit}
\end{figure}

\subsection{Best-Fit Model}
\label{sec:bestfit}

The parameters of the best-fit model and good-fitting models are summarized in Table \ref{tab:parameters}. Our best-fit model has an inner edge at \rinbest\,$R_\text{WD}$ and relatively narrow radial width of \deltarbest\,$R_\text{WD}$. The half-opening angle is \halfanglebest$^\circ$. The dust mass is \massbest\,$\times$\,$10^{18}$ g and the minimum grain size is \aminbest\,$\micron$.

\begin{figure*}
\epsscale{1.17}
\plotone{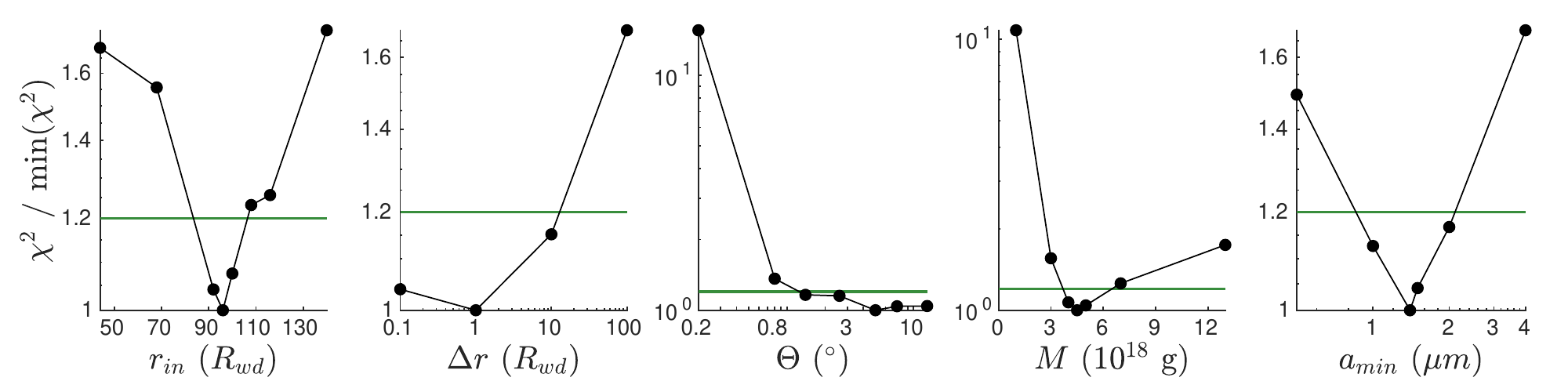}
\caption{Lowest $\chi^2$ value found in the grid for each value of each parameter, normalized to the overall minimum $\chi^2$. The green line indicates the threshold of $\chisqthresh$ times the minimum $\chi^2$, below which models are considered to be good fits.}
\label{fig:chisqcontours}
\end{figure*}

The spectra of all $\numgoodmodels$ good-fitting models from the grid are plotted in Figure \ref{fig:bestfit}. While the models do not match the data perfectly in all wavelength ranges, they do capture the most salient features of the IR spectrum: excess starting at $\sim$1.7 $\micron$, a featureless excess peaking at $\sim$5 $\micron$, a prominent silicate feature at $\sim$10 $\micron$, a shoulder at $\sim$18 $\micron$, and a steep decrease in flux toward longer wavelengths. This is the first time that all these aspects of the IR spectra have been reproduced from a single coherent physical disk model. 

The most prominent discrepancy is around the silicate feature, with the model feature being broader than the data. This could be due to our use of Mie theory, which assumes grains are perfect spheres. Model opacity spectra of non-spherical grains show differences in their IR solid state features and generally better agreement with laboratory measurements \citep[e.g.,][]{mutschke2009_grainshape}. Another discrepancy is that the J and K photometry are fainter than the photosphere model, likely due to variability from stellar pulsations. This has no impact on the fit, as the disk contributes little flux at these wavelengths.  

\begin{figure*}
\epsscale{1.17}
\plotone{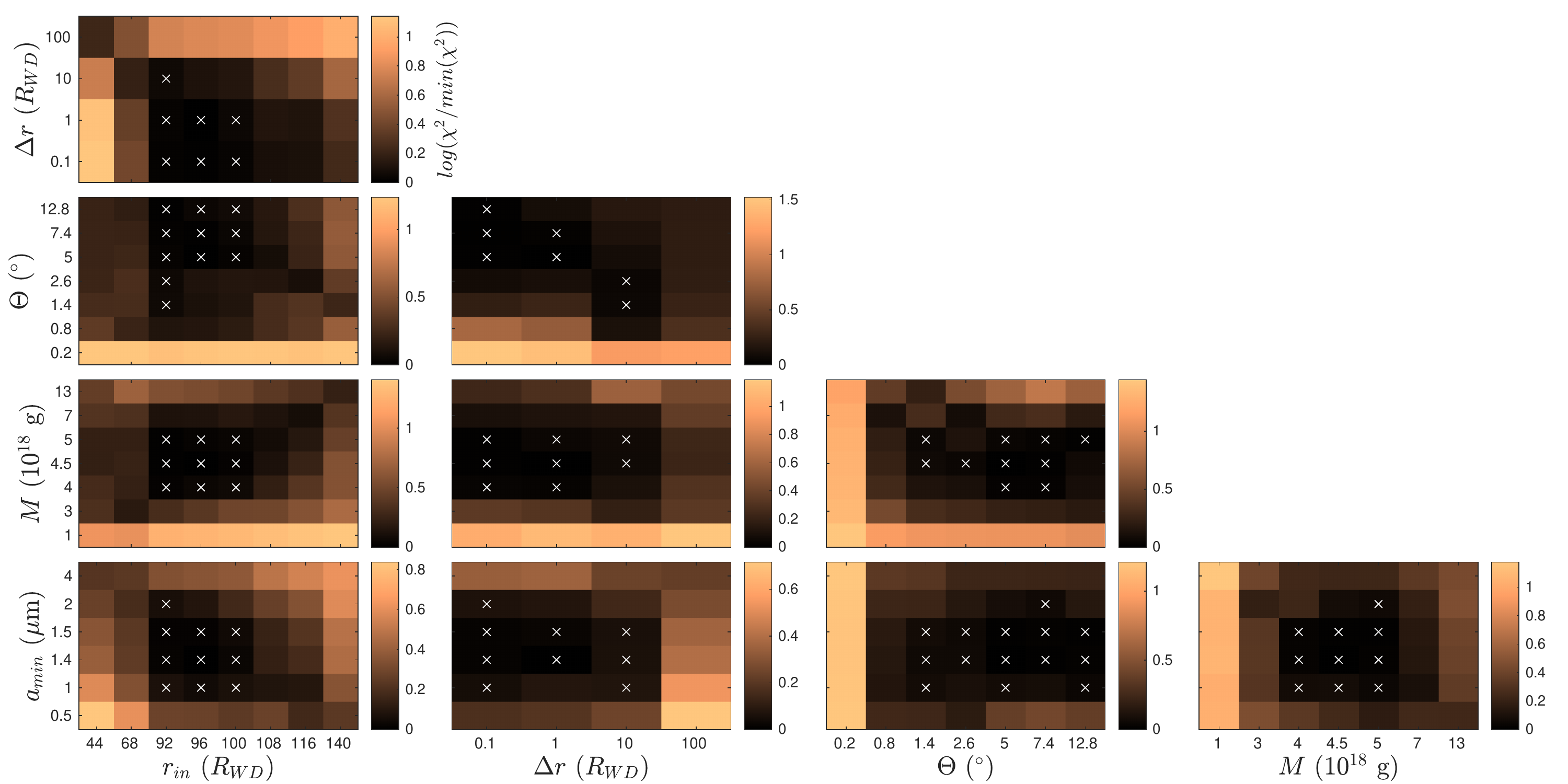}
\caption{Lowest $\chi^2$ value found for each pair of parameter values in the model grid. Color corresponds to the base-10 log of the $\chi^2$ normalized to the minimum $\chi^2$ value. Parameter combinations marked with an $\times$ have good-fitting models with $\chi^2$ $<$ $\chisqthresh$ times the minimum $\chi^2$. Note that the parameter values are not uniformly spaced.}
\label{fig:2Dchisq}
\end{figure*}

In Figure \ref{fig:chisqcontours} we show the 1D $\chi^2$ contours for each parameter---that is, the best $\chi^2$ value achieved in the grid with each parameter fixed at each value. We show the 2D version of these constraints in Figure \ref{fig:2Dchisq}, with parameter pairs that achieve a good fit marked with an $\times$.   

\subsection{Impact of Each Parameter}

\begin{figure*}
\epsscale{1.07}
\plotone{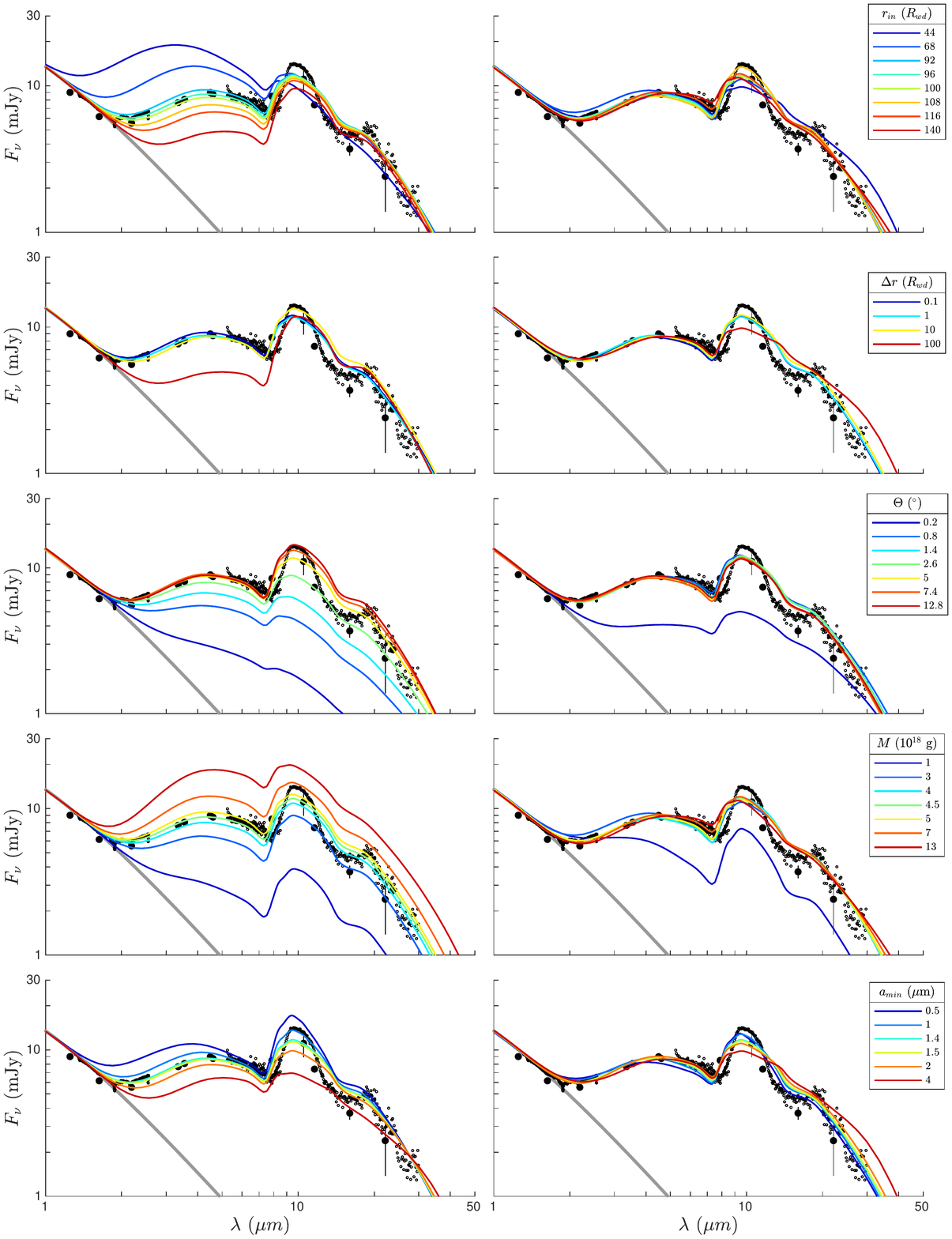}
\caption{Left: effect on the model spectra by varying each parameter individually while keeping the other parameters fixed at the values of the overall best-fit model. Right: models when a single parameter is fixed while allowing the other four parameters to converge to their best-fitting values. The legend shown in the right panel also applies to the models shown in the left panel for each row.}
\label{fig:parameterSEDs}
\end{figure*}

We illustrate the effect of each parameter on the model spectra in Figure \ref{fig:parameterSEDs}. In the left column, one parameter at a time is varied while holding all other parameters fixed at the overall best-fit values. The right column shows models where one parameter is fixed while the other four parameters are free to converge to the best-fitting model. This illustrates how degeneracies among the parameters can lead to good fits for a range of parameter values, as well as how models outside of this range are excluded by their poor fits.

The inner edge location strongly affects the shorter wavelengths of the model. Smaller values place the dust closer to the star, leading to a brighter excess arising at shorter wavelengths.    

Wider belts (with the same inner edge location and total mass) shift more mass to larger radii, resulting in the model being fainter at short wavelengths. Compensating for this in the fitting results in models that are too bright at longer wavelengths. The smallest $\Delta r$ value in our model grid (0.1 $R_\text{WD}$) yields a good fit, so our constraints on $\Delta  r$ are best described as an upper limit.

The half-opening angle determines the size of the disk as seen by the star and thus the amount of radiation intercepted. Higher/lower values increase/decrease the flux at all wavelengths. Our fitting results in a lower limit on $\Theta$.

The total dust mass also increases or decreases the model flux at all wavelengths. Lower masses result in a more prominent silicate feature relative to the continuum, as more of the disk is optically thin. In contrast, high masses have higher optical depths, resulting in a muted silicate feature.

Smaller grain sizes lead to a brighter continuum flux, especially at short wavelengths, because there is more grain surface area for the same total mass, and because small grains achieve a higher temperature than larger grains in the same radiation field. Smaller grains also result in a more prominent silicate feature. Larger values of $a_\text{min}$ can be compensated for in the fitting to some extent with smaller values of $r_\text{in}$, and vice versa. However, the largest value of $a_\text{min}$ considered (4 $\micron$) results in a very broad silicate feature and flux at 20--30 $\micron$ that does not decrease with wavelength as steeply as the data.

\subsection{Dust temperature}
\label{sec:dusttemp}

\begin{figure*}
\epsscale{1.17}
\plotone{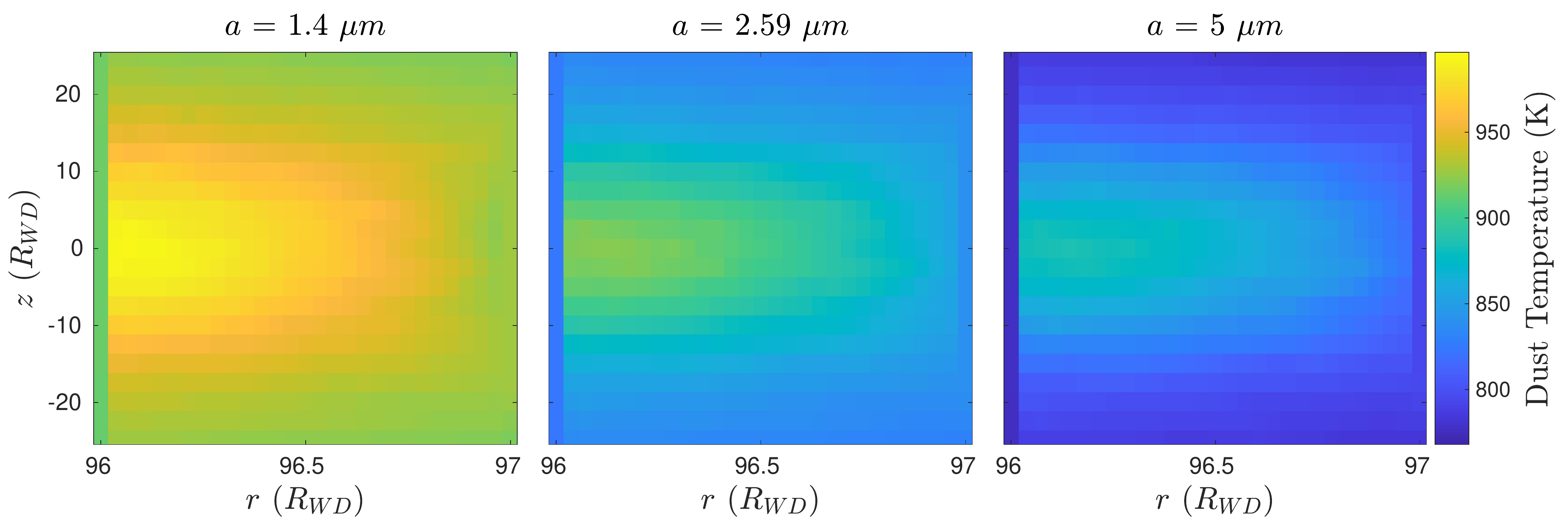}
\caption{Dust temperature distribution of the best-fitting model at three different grain sizes. The pixels in these images, which are linearly spaced in cylindrical coordinate, do not correspond to radiative transfer model cells, which are non-linearly spaced and in spherical coordinates. We plot the temperature up to $z$ = $\pm 3 H_\text{out}$. Note that the axis dimensions are not to scale.}
\label{fig:dusttemp}
\end{figure*}

The radiative transfer calculation computes the dust temperature throughout the disk for each grain size in the model. In Figure \ref{fig:dusttemp} we show the $z$ versus $r$ temperature distribution of our best-fitting model for three grain sizes. We find that the temperature decreases radially across the disk, as expected, and that small grains are warmer than large grains, also as expected. The maximum grain temperature of $\sim$1000 K is somewhat less than the sublimation temperature of silicates ($\sim$1200 K), although the sublimation temperature depends strongly on the grain chemical composition and internal structure \citep{mann2004_SSdust}, which are not well constrained here. This confirms that our model is physically consistent (i.e. does not require dust hotter than the expected sublimation temperature). However, it also leaves open the question of what, if not sublimation, sets the inner boundary of the dust distribution.

\section{Discussion}
\label{sec:discussion}

\subsection{Minimum Grain Size}
\label{sec:minsize}

Our model's minimum grain size of \amingood\, $\micron$ is larger than the value of 0.1 $\micron$ used in the models of \citet{reach2009} and \citet{xu2018}. Radiation pressure is very weak around white dwarf stars, so the minimum grain size is not set by the ``blowout" size as in debris disks around many main-sequence stars.\footnote{Radiation pressure is insignificant for grains on circular orbits, as we assume for our model. Grains on very eccentric orbits around a white dwarf are susceptible to radiation pressure \citep{brouwers2022_WDdisk}.} \citet{kenyon2017_WDcollisions} note that small grains are likely to sublimate rapidly and predict a minimum grain size of 0.1--1 $\micron$. 

The dust clouds that transit WD 1145+017 exhibit wavelength-independent extinction out to 4.5 $\micron$, indicating the grains must be larger than 1.5 $\micron$ in size \citep{xu2018_transits}. \citet{xu2018_transits} argue that smaller grains are rapidly sublimated. This is similar to the range of values found by our set of models for the minimum grain size for the G29-38 disk. Thus, our results are consistent with a minimum grain size set by sublimation although, as shown in Section \ref{sec:dusttemp}, our model dust temperature does not quite reach the expected sublimation temperature for silicates.

\subsection{Disk Mass}
\label{sec:masssize}

Prior models that assumed a fully optically thick disk are insensitive to the dust mass, so we cannot compare our results with such models. When comparing with optically thin models, we must be aware of the assumed maximum grain size, as most of the mass will be in the largest grains. The optically thin model of \citet{reach2009} had 2 $\times$ $10^{19}$ g in grains up to $a_\text{max}$ = 10 $\micron$ (1.6 $\times$ $10^{19}$ g in grains up to 5 $\micron$ for their power-law distribution). \citet{xu2018} found 4.8 $\times$ $10^{18}$ g in grains up to 3 $\micron$ (6.2 $\times$ $10^{18}$ g in grains up to 5 $\micron$) in their optically thin astronomical silicate model. This does not include the forsterite or optically thick components of their model. Overall, our derived dust mass of \massgood\,$\times$ $10^{18}$ g is somewhat lower than found by previous models but consistent within uncertainties.

A recent measure of the X-ray flux from G29-38 derived an instantaneous accretion rate of 1.63 $\times$ 10$^9$ g s$^{-1}$ of material onto the white dwarf \citep{cunningham2022}. At this rate, a disk with our best-fit mass of \massbest\,$\times$ $10^{18}$ g would have a lifetime of \lifetimesec\, s (\lifetimeyr\, yr). This lifetime is only a few times longer than the time since this disk was discovered in 1987, and no substantial mass depletion has been observed. (Note that this lifetime is a lower limit, as our mass measurement only includes grains up to 5 $\micron$.)  Instead, the IR excess at 10 $\micron$ was observed to increase by 10\% from 2004 to 2007, likely due to an increase in dust mass \citep{xu2018}.

White dwarf disks can be replenished before they are fully depleted. This could be from a single large planetesimal that continuously sheds mass, such as the actively disintegrating body around WD 1145+017 \citep{vanderburg2015_transits} or the remnant planetesimal inferred around SDSS J1228+1040 \citep{manser2019_WDgas}. Alternatively, the disk could be resupplied by a continuous infall of small planetesimals from larger orbital distances. This picture is supported by the widespread IR variability of white dwarf disks \citep{swan2020_WDvariability} as well as by a statistical analysis of white dwarf accretion rates inferred from their atmospheric pollution \citep{wyatt2014_WDaccretion}.

The evolution of white dwarf dust disks was modeled numerically by \citet{kenyon2017_WDcollisions}. They find that collisions play a dominant role in disk evolution, breaking larger solids down into small dust grains, which are then destroyed by sublimation. This differs from the traditional flat-disk model where destructive collisions have a negligible impact on disk evolution, and Poynting-Robertson (P-R) drag drives mass loss as grains drift inward. 

In collisional models where no new material is added to the disk, the disk mass drops by several orders of magnitude after $10^2$--$10^4$ yr, with the IR excess falling below detectable levels after 10--100 yr \citep{kenyon2017_WDcollisions}. This timescale does not strictly rule out this scenario for the G29-38 disk, but it would require the disk to have formed not long before it was discovered.

The scenario of ongoing disk replenishment was also modeled by \citet{kenyon2017_WDcollisions}, who found that the disk mass reaches a stable equilibrium. This equilibrium mass (their Equation 11) depends on the white dwarf mass, mass input rate, size of the infalling planetesimals, material density, and the disk radial location, width, and eccentricity. Using the X-ray-measured accretion rate for the mass input rate, 1 km for the planetesimal size, 3 g cm$^{-3}$ for the material density, 0.01 for the disk eccentricity, and our best-fit model for the disk location and width, we find an equilibrium mass of \Mdeq\,g. This mass is within a factor of $\sim$5 compared to the disk mass of our best-fit model, which is a fairly good agreement considering the inherent uncertainties in many inputs to this calculation.

If disks are resupplied by multiple infalling planetesimals, which are unlikely to have identical compositions, we might expect the composition of the white dwarf pollution to vary over time. Yet, the equivalent widths of the metallic lines in the G29-38 atmosphere appear stable \citep{debes2008_G2938stability}. Similarly, the pollution of the white dwarf GD 56 was observed to be stable over a 15 yr timescale despite its IR excess varying by $\sim$20\% \citep{farihi2018_GD56}. This suggests disk and accretion processes are decoupled to some degree. A detailed investigation of the implications of our disk models on the accretion process is beyond the scope of this paper.

\subsection{Vertical Structure}

Prior models assumed the disk was vertically flat (and optically thick) or optically thin with the vertical structure unconstrained, leaving little ground for quantitative comparison. The scale height at the inner edge of our model is $H_\text{in}$ \Hinmin~$R_\text{WD}$. This is larger than the stellar radius, so the disk is primarily illuminated at its front edge, rather than at its upper and lower surfaces. Our model thus corresponds to the ``physically thick" regime discussed by \cite{reach2009}. In this regime, the half-opening angle must be greater than 0.8$^\circ$ in order to intercept enough stellar radiation to account for the total disk luminosity \citep{reach2009}. Our models, with $\Theta$\,\halfanglegood$^\circ$, are consistent with this constraint.

Flat disk models were motivated by the idea that grains should dynamically settle \citep{jura2003}. The disk evolution model of \citet{kenyon2017_WDcollisions} found that the collisional timescale is shorter than the settling timescale. That is, planetesimals are broken down into dust and lost via sublimation faster than those particles can settle. Our set of models are consistent with the idea that significant vertical settling has not taken place, which would support the conclusion that collisions play a major role in white dwarf disk evolution.

Ongoing infall of small planetesimals onto the disk, as discussed in Section \ref{sec:masssize}, could maintain its vertically extended structure. In the WD 1145+017 system, clouds of dust released from actively disintegrating planetesimals transit in front of the star, but the existing disk of dust does not \citep{xu2018_transits}. This demonstrates that new material can enter a disk at a different inclination and thus contribute to the disk's scale height.

With collisions playing a dominant role in the G29-38 disk, we can compare its vertical structure with that of other systems dominated by collisional evolution: main-sequence debris disks. It is most salient to compare with observations of debris disks in the (sub)millimeter, which trace the large grains that are less affected by radiation pressure and gas in these disks \citep{olofsson2022_DDvertical}. 

The edge-on Kuiper belt-like debris disk around $\beta$ Pictoris has dust in two overlapping vertical distributions: a dynamically cold component with a half-opening angle of $\Theta$ = 0.8$^\circ$ and a dynamically hot component with $\Theta$ = 6.3$^\circ$, as measured by ALMA \citep{matra2019_betapic}. The AU Mic debris disk was imaged with ALMA to have a half-opening angle of $\Theta$ = 1.1--1.8$^\circ$, depending on the wavelength and specific analysis \citep{daley2019_AUMic,vizgan2022_AUmic}. Similar analyses found $\Theta$ = 2.3$^\circ$ for the disk around HR 4796 \citep{kennedy2018_HR4796} and $\Theta$ = 2.9$^\circ$ for the disk around q$^1$ Eri \citep{lovell2021_q1Eri}. These values are consistent with (or slightly lower than) the half-opening angle values our models obtain for the G29-38 disk, which would suggest this system is more similar to main-sequence debris disks than to the rings of Saturn.

\subsection{Radial Location and Width}

The location of the disk inner edge in our model (\ringood\,$R_\text{WD}$) is significantly farther from the star than in previous models, and the radial width (\deltargood\,$R_\text{WD}$) is smaller. For comparison, the flat opaque disk model of \citet{jura2003} had $r_\text{in}$ = 11 $R_\text{WD}$ and $r_\text{out}$ = 31--69 $R_\text{WD}$, and \citet{vonhippel2007} found similar values. The optically thick component from the model by \cite{xu2018} spans 8--25 $R_\text{WD}$, while their optically thin component spans 8--834 $R_\text{WD}$. The best fitting ``moderately optically thin" model of \citet{reach2009} had its inner edge at 50 $R_\text{WD}$. The more distant inner edge inferred from our model is likely due to its vertical height. That is, our model has a front edge that can be directly illuminated and heated by the star, whereas flat disk models are only illuminated on their top and bottom surfaces at a shallower angle. \citet{vonhippel2007} noted that the absence of a directly illuminated front edge in the flat disk model would lead to an underestimate of the inner edge location.

While no gas is detected in G29-38, the larger $r_\text{in}$ compared to flat disk models has implications for the gas versus dust location in systems that do possess both. In such systems, the dust disk location has been inferred from optically thick models, and the gas and dust have been found to be radially co-located \citep{brinkworth2009_WDdust,melis2010_WDdisks}. This finding is inconsistent with the hypothesis that the gas disk originates at the inner edge of the dust disk via sublimation and extends inwards from there to the stellar surface. Rather, it supports the idea that gas is produced throughout the extent of the dust disk by sublimation or grain sputtering \citep{jura2008_WDsmallasteroids}. Revisiting such systems by fitting their dust disks with radiative transfer models, as we have done for G29-38, may change the picture of the relative locations of gas and dust and perhaps alter our understanding of the origin of white dwarf gas disks. Unfortunately, at the moment, none of the white dwarfs hosting gas disks have mid-IR spectra of their dust disks, which are essential for fitting with radiative transfer models.   

What might cause the G29-38 dust disk to be radially narrow? One possibility is that the tidal disruption radius (the orbital location within which planetesimals are destroyed by tidal forces), which may set the disk outer edge, and the sublimation radius, which may set the inner edge, happen to be close together for this system. The tidal disruption radius for G29-38 is $\sim$115 $R_\text{WD}$ \citep{jura2003}, which is near to the dust location of our model. As discussed previously, the temperature of the small grains at the inner edge is close to, but somewhat below, the expected sublimation temperature. Another possibility is that the dust is produced only in a narrow ring, perhaps just inside of the tidal disruption radius, and fails to spread radially. P-R drag is the primary mechanism by which dust grains could drift inwards, but in white dwarf disks the P-R drag timescale is longer than the collision timescale \citep{farihi2008_IRAC1,kenyon2017_WDcollisions,farihi2018_GD56}. Thus, grains are destroyed by collisions (and sublimation) before they can spread radially. The observed variability in the G29-38 excess can be explained by a change in dust mass without varying its location \citep{xu2018}. Similarly, the variability in the GD 56 excess is color-independent \citep{farihi2018_GD56}, implying the dust location is  constant. This suggests that while some aspects of a white dwarf dust disk are variable, there is some process that stabilizes its location.

\subsection{Model Limitations}
\label{sec:limitations}

Our model assumes an axisymmetric disk with a grain size distribution independent of location. Theoretical work suggests that disks around white dwarfs can be eccentric, especially during their initial formation from a tidally disrupted asteroid \citep{nixon2020_WDdisk,brouwers2022_WDdisk}. While our results show that the IR excess can be reproduced with a circular disk, they do not rule out disks with inherently different geometries. We suggest that radiative transfer be used to calculate the IR spectra of models derived from detailed simulations of disk formation and evolution.

In our fitting, we fixed $\gamma$ = 1. Because a radially narrow disk configuration is favored, $\gamma$ does not have a large effect on the model. We also fixed $i$ = 30$^\circ$ in the fitting. Exploring the effect of inclination on our best-fit model, we find the SED to be significantly fainter for $i$ = 0$^\circ$--20$^\circ$ because the illuminated front edge of the disk is less visible when viewed top-down. For $i$ = 30$^\circ$--80$^\circ$ (the most likely range of inclinations), the brightness increases somewhat with increasing inclination, but the effect is not large. At $i$ = 90$^\circ$, the disk partially occults itself and the central star, so the disk is fainter, especially at shorter wavelengths. 

Lastly, we fixed $a_\text{max}$ = 5 $\micron$. To study this assumption, we make a small number of additional models with $a_\text{max}$ = 10 $\micron$. This larger maximum grain size leads to a somewhat fainter excess, as more of the mass now resides in larger grains. We are able to recover a decent fit to the data by also increasing $M$ to 5$\times$10$^{18}$ g, decreasing $r_\text{in}$ to 92 $R_\text{WD}$, and decreasing $a_\text{min}$ to 1 $\micron$. These values fall within the range of good fits from our original grid, so we conclude that our choice of $a_\text{max}$ does not significantly bias the fit.       

\section{Summary}
\label{sec:summary}

We fit the IR measurements of the G29-38 white dwarf dust disk for the first time using radiative transfer models based on a coherent disk structure. Prior modeling efforts had employed an optically thick (vertically flat) model, a purely optically thin model, or a linear combination of the two. We explore a grid of models with five free parameters: the inner edge location, the radial width, the vertical half-opening angle, the total dust mass, and the minimum grain size.

We find that the disk mass (\massgood\,$\times$ 10$^{18}$ g) is generally consistent with (or somewhat lower than) prior models. The minimum grain size (\amingood\,$\micron$) is somewhat larger than in prior models, but consistent with theoretical expectations for dust sublimation. The half-opening angle (\halfanglegood$^\circ$) is inconsistent with the flat disk model and in line with the vertical structure seen in debris disks around main-sequence stars. The inner edge location (\ringood\,$R_\text{WD}$) is significantly farther from the star than found by prior models because those models did not include a directly illuminated inner edge. The radial width (\deltargood\,$R_\text{WD}$) is narrower than found by prior models, perhaps due to inefficient spreading or the proximity of the tidal disruption radius to the sublimation radius.

Overall, our models would be consistent with the growing picture of white dwarf disks as collisionally active and continuously fed by fresh material. The smallest grains are ultimately destroyed by sublimation. Short collisional timescales can explain why the dust does not settle vertically or spread radially. Thus, within our model framework, white dwarf disks would be similar to main-sequence debris disks.

\medskip

We thank the referee for providing helpful feedback on this paper. We thank Dr. Mukremin Kilic for sharing with us the SpeX data. We thank Siyi Xu, Amy Bonsor, Erik Dennihy, and Johan Olofsson for useful feedback on this project. We also gratefully acknowledge the use of the University of Virginia Rivanna High-Performance Computing system, which was utilized to run the radiative transfer models presented in this work. N.P.B and K.Y.L.S acknowledge support from NASA XRP grant 80NSSC22K0234. N.P.B. and C.I.L. acknowledge support from the Virginia Initiative on Cosmic Origins (VICO).

\software{RADMC-3D \citep{dullemond2012}}

\bibliographystyle{aasjournal}
\bibliography{NPB}

\end{document}